\documentclass{article}[15pt]

\usepackage[dvips]{epsfig}
\usepackage{rotating}

\begin{document}

\title{
Extreme Value Distribution Based Gene Selection Criteria for 
Discriminant Microarray Data Analysis Using Logistic Regression
\vspace{0.2in}
\author{
Wentian Li$^{1,*}$, Fengzhu Sun$^2$, Ivo Grosse$^3$ \\
{\small \sl 1. The Robert S Boas Center for Genomics and Human Genetics,
North Shore LIJ Research Institute,}\\
{\small \sl 350 Community Drive, Manhasset, NY 11030. USA.}\\
{\small \sl 2.  Department of Biological Sciences, Molecular and Computational Biology Program,}\\
{\small \sl University of Southern California, 
1042 W. 36th Place, DRB 155, Los Angeles, CA 90089. USA.}\\
{\small \sl  3. Institute for Plant Genetics and Crop Plant Research, Corrensstrasse 3, D-06466 Gatersleben. Germany.} \\
{\small \sl * Corresponding author.}
}
\date{To be published in {\sl Journal of Computational Biology}, Vol.11, Nos.2-3 (2004)}
}
\maketitle    
\markboth{\sl Li,Sun,Grosse}{\sl Li,Sun,Grosse}

\newpage

\large

\begin{center}
{\bf Abstract}
\end{center}

One important issue commonly encountered in the analysis of
microarray data is to decide which and how many genes should be
selected for further studies.  For discriminant microarray data
analyses based on statistical models, such as the logistic
regression models, gene selection can be accomplished by a
comparison of the maximum likelihood of the model given the real
data, $\hat{L}(D|M)$, and the expected maximum likelihood of the model
given an ensemble of surrogate data with randomly permuted label, $\hat{L}(D_0|M)$.  Typically, the
computational burden for obtaining $\hat{L}(D_0|M)$ is immense, often
exceeding the limits of computing available resources by orders of magnitude.
Here, we propose an approach that circumvents such heavy
computations by mapping the simulation problem to an extreme-value 
problem. We present the derivation of an asymptotic distribution of the extreme-value
as well as its mean, median,  and variance.  Using this distribution, 
we propose two gene selection criteria, and we apply them to two
microarray datasets and three classification tasks for illustration.

\vspace{0.8in}

{\bf Key words:} microarray, gene selection, extreme value distribution,
logistic regression

\newpage

\section{Introduction}

\indent

Discriminant microarray data analysis can be understood as a
comparison of the expression levels of samples from one 
group versus another group, such as disease tissues 
versus normal tissues, or one subtype of cancer versus another 
subtype (for a review, see \cite{speed}). Discriminant analysis or 
classification can be carried out on a whole set of genes or 
on individual genes, and it has become increasingly clear 
that, for many classification tasks based on microarray data, 
it is not necessary to consider many genes simultaneously.  
In many cases it has been shown that a few genes are sufficient 
for classifying two groups of samples
\cite{anto,fc,ding,jae,lee,lihong,jli,liyang1,lu,model,nguyen,xiong}. 
Usually, even with a very small number of genes being included
in a classification, these genes are jointly used in a
multivariate fashion. However, in some cases, one or two 
genes are sufficient for a good classification \cite{liyang1,siedow,xiong}. 
This observation led to procedures that examine one gene at a time,
rank the gene according to their classification ability, and select 
only the high-ranking genes for further studies, including new
confirmation experiments \cite{broberg,marta,smyth}. Some information
could be lost by not considering genes jointly, but focusing on
single genes often simplifies the biological interpretation of
the results.

Two single-gene classification methods that are often applied to the
analysis of microarray data are the fold-change method \cite{chen} and 
the $t$-test \cite{sokal}.  As repeatedly pointed out in 
Refs.~\cite{baldi,claverie,draghici,ideker,mutch,pan,thomas},
the fold-change method is not rigorous from a statistical point of
view, because it considers neither the variances nor the sample sizes of the
data. For example, a two-fold increase obtained from  narrowly
distributed data with 1000 samples is statistically more
significant than the same increase obtained from broadly distributed
data with 10 samples.  The $t$-test overcomes this shortcoming by
including the variance and sample size information. However, the $t$-distribution is obtained by
assuming that the random variables are sampled from a normal (Gaussian)
distribution.

There are alternative discriminant methods that do not rely
on the assumption that the random variables are normally 
distributed.  Out of the four linear classification methods -- Fisher's linear 
discriminant analysis, logistic regression (LR), 
Rosenblatt's perceptron, and support vector machine (SVM) -- 
LR and SVM do not rely on this assumption \cite{hastie}, and 
hence they are more
robust when the actual data, including the presence of outliers,
are not normally distributed. Another advantage of LR over $t$-tests 
is that $t$-tests compare only two group averages, whereas LRs
check each individual sample for consistent differential expressions. 
In the following we focus on LR, which has already been used 
in discriminant microarray data analyses 
\cite{eilers,liyang1,liyang2,nguyen,shevade,vant}.

Cross-validation is often used for assessing  how accurately 
a dataset can be classified by a learned model.
In cross-validation, a dataset is divided into two parts, 
where the first part is used for estimating the model parameters, 
and the second part is used for assessing the classification performance. 
Due to the splitting of the dataset, not all samples are 
included in the learning process, which is not optimal for 
datasets with a small number of samples. On the other hand, if
all data points are used in the training process, the error 
rate of the classifier would be underestimated.

In order to estimate the statistical significance of a
learned model, one usually uses resampling methods,
such as the bootstrap method (resampling with replacement)
or the  permutation method (resampling without replacement). 
Since in this paper only the single-gene LR is used, a 
significant model implies a significant gene. (This correspondence
does not hold for multivariate classifiers due to the possible
correlation among genes.) In Ref.\cite{liyang2},  likelihoods of
single-gene LRs of real datasets are compared to those of the
label-permuted datasets, and genes with a likelihood
exceeding the likelihood of the top-ranking gene of
the permuted data are selected. One problem with actually carrying
out permutations as in Ref.\cite{liyang2} is that the calculation of 
the LR likelihoods for ten-thousands of genes is computationally 
intensive, and that repeating this calculation for, say, 
$10^4$ surrogate datasets is prohibitive.

Here, we propose an analytic solution that circumvents these
heavy computations.  Our approach is based on the observation
that we are only interested in the extreme-values in the following
sense: in order to define a threshold for gene selection, we compare
the maximum likelihood of each gene in the real data with the maximum
likelihood of the {\sl top-ranking gene} in the label-permuted
data.  Whereas simulation requires the calculation 
of all single-gene likelihoods in the surrogate data  for each
permutation, the proposed analytic calculation of the
the expected value of the likelihood of the top-ranking gene
will be carried out only once.

The extreme-value theory is a well studied topic in statistics
\cite{coles,gumbel-book,resnick}, with major contributions
by Ronald A Fisher,  Maurice Frechet, Emil Gumbel, Vilfredo Pareto, Waloddi
Weibull, to name just a few. One fundamental assumption often used in
deriving an extreme-value distribution is that observations are
independent. In our application of the extreme-value distribution,
the corresponding assumption is that log likelihood scores of different
genes are statistically independent. Clearly, this assumption is 
violated in most expression data sets, but as we discuss in 
the Discussion section, there is a simple solution to this problem 
by replacing the number of genes $p$ by the ``effective number of genes" $p_{\rm eff}$.

The topic studied in this paper is closely related to the
multiple testing problem. A criterion for claiming statistical
significance should be more stringent when many genes are
tested than if only one gene is tested, because presumably multiple testings
provide more chances to find a significant gene. Traditionally,
the Bonferroni correction, which divides the
threshold for significance obtained from a single gene by
the total number of tests (genes), is used in those cases. Applying extreme-value distribution
achieves a similar goal because the largest value among $p$
variables increases with $p$, and this effectively raises
the stringency for a gene selection criterion.

\section{Methods}

\subsection{Logistic regression of microarray data}

\indent

First, we introduce the following notation.  Let the samples be
indexed by $i$, and let the genes be indexed by $j$.  Denote the total
number of samples by $N$, the total number of genes by $p$,
the expression level by $x$, e.g., $x_{ij} = {\rm log}$(spot 
intensity of gene $j$ in sample $i$), and
the sample label value by $y$, e.g., $y=0$ or $y=1$ for a binary
classification problem.  Then, the single-gene LR model $M_j$ of gene
$j$ is defined by the conditional probabilities of the sample label 
$y_i$ given the expression levels $x_{ij}$,
\begin{equation}
\label{eq_lr}
\hspace{0.2in} \Pr ( y_i=1 | x_{ij}) = \frac{1}{1+ e^{-a_j -b_j x_{ij} }},
\end{equation}
for $i=1,2, \ldots, N$ and $j=1,2, \ldots, p$.  Here, $a_j$ and $b_j$
are parameters to be estimated from all samples $i=1,2, \ldots, N$.
The data-fitting performance of $M_j$ is measured by the
maximum likelihood,
\begin{equation}
\label{eq_like_lr}
\hat{L}(D|M_j) = \max_{a_j, b_j} \prod_{i=1}^N
[ \Pr (y_i=1 | x_{ij})]^{y_i} [1- \Pr (y_i=1|x_{ij})]^{1-y_i}, 
\end{equation}
where $D$ denotes the data.
Since a gene is represented by a LR model, selection of genes 
becomes selection of single-gene LR models with large
maximum-likelihoods. Although in a more general context such
as multivariate models, model selection is not equivalent 
to variable (gene) selection, for single-gene models, gene
selection and model selection are treated as the same.

\subsection {Maximum likelihood for the surrogate data}

\indent

There are different ways of constructing surrogate datasets.
For example, one may sample the expression levels $x_{ij}$ from 
a normal distribution, and then assign a label $y_i$ to each sample
randomly; or one may start with the available microarray data 
set, and randomly permute the sample label.  If a gene in 
the microarray data does not differentially express before
a permutation, the two ways for generating the surrogate data 
is the same. However, as pointed out in \cite{pan03}, if a gene is indeed 
differentially expressed before a permutation, extra variance remains
after permutation, and the two methods for generating the surrogate data
can be slightly different.

This subtle difference between the two surrogate datasets may affect
a $t$-test result, because $t$-test makes certain assumption
on the distribution and variance on the data \cite{pan03}. The 
extra variance remained in the permuted data violates this assumption.
Nevertheless, no such assumption is required for LR. For this
reason, we do not make this distinction, and denote by $D_0$ a 
surrogate dataset with permuted sample labels, whether the original 
dataset before permutation contains differentially expressed genes or not.

We denote by $\hat{L}(D_0|M_j)$ the maximum likelihood under the single-gene LR
model $M_j$.  For a particular permutation, we define by
\begin{displaymath}
l \equiv \max_j [ \log \hat{L}(D_0|M_j) ]
\end{displaymath}
the maximum value of the maximum likelihoods of all genes.  Note
the two different maximization steps: the first over the parameter 
values $a_j$ and $b_j$ for a given gene, and the second over all genes $j$.
When surrogate dataset  $D_0$ is repeatedly generated, those maximum values 
$l$ vary from realization to realization, and our goal is 
to characterize the distribution of $l$, e.g.~by computing 
the expected value, the median, or the standard deviation of $l$.

Toward the calculation of the expected value of $l$, we use the
Wilks theorem \cite{wilks}, which is ``one of the most celebrated folklores
in statistics" \cite{fan} and is covered by most standard textbooks
on mathematical statistics
\cite{casella,cox,ferguson,kendall,wilks-book}.  This theorem states that,
under very general conditions (which our LR model
satisfies), the asymptotic distribution of the 2-log-likelihood ratio
-- when the data is generated by the null model $M_0$ -- is 
the $\chi^2$ distribution with $df$ degrees of freedom, 
where $df = d(M_j)-d(M_0)$ is the difference of the number of
parameters in models $M$ and $M_0$ \cite{wilks}.
Using our notation, it states that in the $N \to \infty$ limit,
\begin{equation}
\label{eq_lratio}
2 \log \hat{L}(D_0|M_j)  =  2 \log  \hat{L}(D_0|M_0) + t, 
\end{equation}
where $t$  denotes a random variable sampled from a $\chi^2$
distribution with $df$ degrees of freedom.

We choose the null model $M_0$ to be the same for all genes, i.e.,
$\Pr (y_i=1 | x_{ij}) = c$ for all $j=1,2, \ldots, p$.  The maximum
likelihood estimate of $c$ is simply the percentage of samples that are
labeled as 1, i.e., $\hat{c} \equiv N_1/N $.  The maximum likelihood
under $M_0$ is
\begin{equation}
\label{eq_like_null}
\hat{L}(D_0|M _0) = \hat{c}^{N_1} (1-\hat{c})^{N-N_1}, 
\end{equation}
and its logarithm is
\begin{displaymath}
\log \hat{L}(D_0|M_0)= -NH
\end{displaymath}
where $H$ is the entropy
\begin{displaymath}
H \equiv -{N_1 \over N} \log {N_1 \over N} - {N-N_1 \over N} \log {N-N_1 \over N}.
\end{displaymath}
Note that $\hat{L}(D|M_0)= \hat{L}(D_0|M_0)$, because the 
percentage $N_1/N$ of samples with sample label $y=1$ is 
the same in $D$ and $D_0$.

Applying the LR model to the surrogate data, we obtain
for the best single-gene maximum log-likelihood (in the large sample
limit $N \to \infty$):
\begin{eqnarray}
l & = & \max_j \left[ \log \hat{L}(D_0|M_j) \right]
  =  \max_j \left[ \log \hat{L}(D_0|M_0) +  \frac{t_j}{2} \right]
        \nonumber \\
& = & -NH + \frac{1}{2} \max \left[  t_1, t_2, \ldots, t_p \right], \nonumber
\end{eqnarray}
where $t_1, t_2, \ldots, t_p$ are $p$ random variables sampled from a
$\chi^2$ distribution with $df$ degrees of freedom. In this example,
$M_j$ contains two parameters, and $M_0$ contains
one parameter, so $df= 2-1=1$.

\subsection {Extreme-value distribution of $\chi^2$-distributed
random variables}

\indent

The extreme-value distribution of normally distributed
random variables has been extensively studied (see, e.g.,
\cite{ferguson}). Gumbel showed  \cite{gumbel,gumbel-lieblein} 
that the extreme-value distribution of the $\chi^2$ 
distributed variables belongs to the same class as that 
of normally distributed variables, which is now called  the standard
Gumbel distribution $exp(-exp(-(x-a)/b))$.  For the case of 
$\chi^2$ distributed variables, the coefficients $a$ and $b$ 
are derived in Ref.~\cite{gupta}. Although this extreme-value 
distribution (of random variables sampled from the $\chi^2$
with one degree of freedom) is known, for the sake of completeness 
we present here a derivation.

Let $t_1, t_2, \ldots t_p$ be statistically independent and
identically distributed (iid) random values from a
$\chi^2$ distribution with one degree of freedom, and define $T_p \equiv
\max[ t_1, t_2, \ldots, t_p]$.  Based on the inequality \cite{casella}:
\begin{equation}
\label{eq_ineq}
\sqrt{\frac{2}{\pi}} \frac{\sqrt{t}}{1 + t} e^{-t/2} \leq 
\Pr (t_i  \geq t)  \leq 
\sqrt{\frac{2}{\pi}} \frac{1}{\sqrt{t}} e^{-t/2}
\end{equation}
and by defining
\begin{displaymath}
c_p \equiv \log \frac{p^2}{\pi \log(p) }
\end{displaymath} 
one finds that for asymptotically large ($ p \to \infty$), the 
cumulative distribution of $ v_p = (T_p-c_p)/2$ 
converges to the double exponential function: \\
\begin{equation}
\label{eq_theorem}
F_v(x) = 
\lim_{p \to \infty} 
F_{v_p} (x) \equiv
\lim_{p \to \infty} 
\Pr \left(
\frac{T_p-c_p}{2} \leq  x \right)  = \exp(-e^{-x}).
\end{equation}

\noindent This result can be derived as follows. For any $x$, we obtain
\begin{displaymath}
\Pr \left(\frac{T_p - c_p}{2} \leq x \right) 
 = \Pr (  T_p \leq c_p + 2x ) 
 = \prod_{i=1}^p \left[ 1- \Pr (t_i > c_p+2x) \right]
 =  \left[ 1 - \Pr ( t_i > c_p + 2x ) \right]^p,
\end{displaymath}
and from inequality (\ref{eq_ineq}), one obtains
\begin{displaymath}
\label{eq_long}
\lim_{p \to \infty} p \Pr ( t_i > c_p+2x)=
\lim_{p \to \infty}
p \sqrt{\frac{2}{\pi} }
\frac{e^{-\log(p)+ \log\sqrt{\log (p)}+ \log (\sqrt{\pi}) -x}}
{\sqrt{c_p+ 2x}}
= \lim_{p \to \infty}
\frac{\sqrt{2 \log (p)}}{\sqrt{c_p+2x}} e^{-x} =e^{-x}
\end{displaymath}
Therefore,
\begin{displaymath}
\lim_{p \to \infty}
\Pr \left( \frac{T_p - c_p}{2} \leq x \right)
= \exp(- e^{-x} )
\end{displaymath}

From the asymptotic distribution $F_v(x)$, we can compute the mean $E[v]$,
the median $m[v]$, and the standard deviation $\sigma[v]$: 
\begin{eqnarray}
E[v] &=& \gamma  \nonumber \\
m[v] &=& -\log (\log(2)) \\
\sigma^2[v] &=& \frac{\pi^2}{6},
\end{eqnarray}
where $\gamma \approx 0.5772$ denotes the Euler constant.
Hence, we obtain the following asymptotic scaling for the mean, the
median, and the standard deviation of $T_p = c_p+ 2 v_p$ in 
the asymptotic limit $p \to \infty$:
\begin{eqnarray}
\label{eq_mean}
E[T_p]   & \approx & 2 \log(p) - \log (\log (p)) - \log (\pi) + 2 \gamma
 \nonumber \\
m[T_p]  & \approx &  2 \log(p) - \log (\log (p)) - \log (\pi) 
-2 \log (\log(2))
 \nonumber \\
\sigma[T_p] & \approx & \sqrt{ 2 \pi^2/3}
\end{eqnarray}

Based on the extreme-value distribution of $T_p$, 
we propose the following  two gene selection criteria.

\subsection{Gene selection based on the E-value of the 
extreme-value distribution}

\indent

In the first criterion, which we call the E-criterion, we
compare the maximum likelihood of each gene obtained from 
the real data with the expected value  of the maximum
likelihood of the top-ranking gene from the surrogate data. 
This criterion for the likelihood can be easily converted
to a criterion for the log-likelihood ratio: for each 
gene $j=1, 2, \ldots p$, calculate the 
log-likelihood ratio
\begin{equation}
\label{eq_tj}
t_j \equiv 2\log \frac{ \hat{L}(D|M_j)}{\hat{L}(D|M_0)}
= 2\log \frac{ \hat{L}(D|M_j)}{\hat{L}(D_0|M_0)}
= 2\log \hat{L}(D|M_j) + 2NH,
\end{equation}
order them such that $t_{(1)} \geq t_{(2)} \geq t_{(3)} \ldots \geq t_{(p)}$,
and declare genes $j=1,2, \ldots J$ as differentially expressed if 
\begin{equation}
\label{eq_criterion1}
t_{(J)} \geq E[T_p]= 2 \log(p) - \log (\log (p)) - \log(\pi) + 2 \gamma > t_{(J+1)}.
\end{equation}

\subsection{Gene selection based on the P-value of the
extreme-value distribution}

\indent

In the second gene selection criterion, which we call the P-criterion,
we compare the $P$-value of the calculated maximum likelihood 
of each gene obtained from the real data using the distribution
of the maximum likelihood of the top-ranking gene from the surrogate data. 
That is, 
for each gene $j=1, 2, \ldots p$, calculate the log-likelihood ratio
$t_j \equiv 2\log \hat{L}(D|M_j) + 2NH$, order them to 
$t_{(j)}$, then convert them to $v_{(j)} \equiv (t_{(j)} - c_p)/2$.
We declare genes $j=1,2, \ldots J$ as differentially expressed 
if and only if an upper limit of the $P$-value for $v_{(J)}$,
$P_{(J)} \le  1 - \exp(-e^{-v_{(J)}})$, is smaller than the 
user-specified $P_0$, and that of $v_{(J+1)}$ is larger:
\begin{equation}
\label{eq_criterion2}
1 - \exp(-e^{-v_{(J)}}) \le P_0 < 1- \exp( -e^{-v_{(J+1)}}).
\end{equation}

When a small $P_0$ is chosen, such as $P_0=$0.01 or $P_0=$0.001, the
tail distribution of the extreme-value is used. In the E-criterion,
since it is the mean of the extreme-value is chosen, we focus
on the middle-range of the extreme value distribution. As a
result, the P-criterion is more stringent than the E-criterion,
leading to fewer genes selected. This is on the top of the
conservative nature of both E- and P-criteria, because even
the non-top genes in the real data are compared with the top-maximum-likelihood 
in the surrogate data.

\section{Results}

\subsection{Confirmation of the extreme-value distribution by numerical
simulation}

\indent

We perform numerical simulations to test if, and to which
degree, the asymptotic expressions of the mean $E[T_p]$, the
median $m[T_p]$, and the standard deviation $\sigma[T_p]$ are
acceptable approximations for finite $p$ ranging from 1 to $10^5$.
For each value of $p$ ranging from $1$ to $1.5 \times 10^5$
we generate $10^4$ samples of $p$ random variables sampled from
a $\chi^2$ distribution with 1 degree of freedom.
Fig.~1 shows $E[T_p]$, $m[T_p]$, and $\sigma[T_p]$ versus $\log(p)$,
and we find that the asymptotic expressions of $E[T_p]$
and $m[T_p]$ agree with the simulation data sufficiently well.
The simulations confirm the trend of a linear increase of $T_p$ with $\log(p)$
as well as the systematic deviation from this
linear trend due to the $\log \log(p)$ term. The standard deviation
$\sigma[T_p]$ according to Eq.(\ref{eq_mean}) is not a function
of $\log(p)$, and indeed, the simulated values reach a plateau for $p>10^3$.
Note that the predicted standard deviation $\sqrt{ 2 \pi^2/3}$
is consistently larger than the simulated standard deviation, and 
the difference between the two curves becomes smaller as $p$ increases.

Besides the mean, median, and variance, we also compare the distribution
of $T_p$ for finite $p$ with the analytically derived distribution for $p \to \infty$ in order to study to which
degree
Eq.(\ref{eq_theorem}) derived for the asymptotic limit $p \to \infty$
is an appropriate approximation for finite p ranging from
$10^3$ to $10^4$.  We generate $p=6000$ random variables
$t_1, t_2, \ldots t_p$ sampled from a $\chi^2$ distribution
with one degree of freedom, and we record the maximum value
$T_p \equiv \max [t_1, t_2, \ldots t_p]$.  We repeat this sampling process $10^4$ times,
and we compare the empirical $P$-value, which
is the percentage of times the $v_p=(T_p-c_p)/2$ exceeds
a specified value $x$, to the theoretical
$P$-value $1-F_v(x)= 1- \exp( -\exp(-x) )$. We find that for $p=6000$ the two distributions
match well.

\subsection{Gene selection for microarray datasets}

\indent

We use two publicly available microarray datasets to illustrate the
proposed criteria for deciding how many high-ranking genes
should be selected: (i) the leukemia subtype data from the
Whitehead Institute \cite{golub}, and (ii) the colon cancer
data from Princeton University \cite{alon}.

{\bf ALL versus AML:} Fig.~2(a) shows
the rank-ordered distribution of the maximum likelihoods for
all single-gene LR models for the discrimination of acute
lymphoblastic leukemia (ALL) from acute myeloid leukemia (AML).
The sample size is 72, which combines both the training and
testing sets, as designated in \cite{golub}. The ALL-AML
classification problem is thoroughly discussed in \cite{lin},
and it is well-known to be a comparatively easy classification
problem \cite{liyang1,lu,nguyen,siedow}.

According to the E-criterion proposed in Eq.(\ref{eq_criterion1}), 407 genes
are selected. In the converted variable $v_{(j)} = (t_{(j)}-c_p)/2$, the
E-criterion is equivalent to $v_{(j)} > \gamma=0.5772$.
Using the P-criterion proposed in  Eq.(\ref{eq_criterion2}),
we obtain that 165 genes are considered to be differentially
expressed at the P-value of 0.01 (see Fig.~2(b)).
We note in passing
that the number of genes selected by both criteria is
substantially smaller than 1100, which is the number of genes
labeled as ``more highly correlated with the AML-ALL class distinction
than would be expected by chance," as reported in \cite{golub}
using the ``neighborhood analysis".

{\bf T-cell versus B-cell:} As pointed out in \cite{grant}, the ALL
dataset is still a heterogeneous dataset, with sources from B-cells
and T-cells being different from each other. Fig.~3(a) shows the
rank-ordered distribution of the maximum likelihoods using
single-gene LR models for the B-cell versus T-cell
classification, with a reduced sample size of 47.  The E-criterion
declares 114 genes as differentially expressed, and
the more conservative P-criterion declares 57 genes as differentially
expressed with the P-value of 0.01.
These findings are in agreement with the observation
in \cite{grant} that there are differentially expressed
genes in B-cells and T-cells, and also in agreement with
another observation in \cite{lli} based on cluster analysis.

{\bf Colon cancer versus  normal:} Fig.~4(a) shows the rank-ordered distribution
of the maximum likelihoods using single-gene LR models for the colon
cancer versus normal tissue dataset studied in \cite{alon}.  This
dataset consists of 62 samples, and the data for 2000 genes that
have the ``highest minimal intensity across the samples" are available from \cite{alon}.
We find that only 49 and 10 genes are selected by the
E-criterion and the P-criterion (at $P_0$-value=0.01), respectively, and one
possible explanation why these numbers are small is that
the initial number of genes is already restricted to a smaller
number of 2000 by some pre-processing method.  Another possible explanation 
is that it the classification task in the dataset cannot be
accomplished by single-gene models.

\section{Discussions and conclusions}

\indent

%

The gene selection procedure discussed here circumvents the multiple
testing problem by explicitly including the number of genes ($p$) 
in the gene selection criterion.  It is an analytic approximations 
based on the known mathematical theorems concerning (i) the 
extreme-value distribution of $\chi^2$ distributed random variables, 
and (ii) the asymptotic distribution of the log-likelihood ratio.
The analytical approximation developed in this paper is based on the 
following assumptions: (1) $N \to \infty$ so that the distribution of 
the log-likelihood ratio statistics is the $\chi^2$ distribution;
(2) $p \to \infty$ so that the extreme-value distribution can be
applied; (3) the extreme-value is taken from $p$ {\sl independent}
values.  In the context of microarray data analysis, these assumptions 
translate to: (1) the number of microarray samples $N$ is very large;
(2) the number of genes $p$ is very large; and (3) the maximum 
likelihood scores of different genes are statistically independent.

Based on the simulation result presented in Fig.~1, problem (2) 
may not be a serious problem, since the $\log(p)$ trend, as well
as the $\log( \log(p))$ correction, is captured very well
by the analytic formula, even when $p$ is small.
Besides,  for a typical microarray data, the range of $p$ is
large, usually beyond a few thousands. It should be mentioned
that any asymptotic results (asymptotic for $p$) are not unique in the
sense that adding any extra term whose value
over $c_p$ tends to zero will also be a valid solution. For example, it can
be shown that it is possible to replace
$c_p= 2\log(p)- \log (\log(p)) + \log(\pi)$
by $c_p' = 2 \log(p) - \log [ \log(p) - \log \sqrt{\log(p)} -\log \sqrt{\pi} ]
 - \log (\pi) )$. At finite range of $p$'s,
however, the difference between different formula can be
neglegible.

Probelm (3) can be handled by introducing an ``effective number
of genes" $p_{\rm eff}$.  For example, if two genes have identical 
expression profiles, they lead to the identical maximum-likelihood 
scores, and the number of genes should be reduced by one, i.e., $p_{eff}=p-1$.
In cDNA arrays, several probes may consist of ESTs originated from 
the same gene, so these probes will give highly correlated expression 
profiles.  Since the exact degree of correlation is usually unknown, 
one must estimate the total number of redundant probes, and subtract 
them from $p$ to obtain $p_{eff}$. As $p_{eff} < p$, 
and $c_{p_{eff}}< c_p$, the effect of a gene-gene correlations 
is to relax the gene selection criterion and hence more genes
are selected.  Interestingly, a few recent publications show
that gene-specific test scores are almost independent \cite{pan03-mixed,tibshirani}.
As a result, the problem (3) may not  be a serious problem
for real data.

When the multiple testing is considered in a $t$-test, 
the gene selection criterion becomes more stringent with
more number of genes. It is the same
situation for the E-criterion and P-criterion. Both E- and P-criterion
are conservative in the sense that the $j$-th ranked gene is
compared to the top-ranked classification performer, instead
of the $j$-th ranked one, in the surrogate data.  If the  E- and 
P-criterion are compared to each other, we find that for 
small values of $P_0$, such as $P_0 = 0.01$ or $P_0 = 0.001$, the 
P-criterion is more stringent than the E-criterion. It is because
the E-criterion uses the average of the extreme-value
distribution whereas the P-criterion uses the tail area of
the distribution.

The conservative nature of the E- and P-criterion yields a side 
effect that fewer number of genes are selected than some other 
gene selection criteria.  This may be a positive or negative 
side effect, depending on the goal of the data analyst.  Selecting many genes 
as differentially expressed increases the risk of declaring 
non-differentially expressed genes as differentially expressed, and 
selecting only a few genes increases the risk of missing differentially 
expressed genes.  In the framework of hypothesis testing, one can 
reduce the type-I error (the number of false positives) at the cost 
of increasing the type-II error (the number of false negatives).  
A too stringent gene criterion reduces the number of false positives 
in the set of selected genes at the cost of missing  potentially
meaningful genes.  Whether or not a good  balance is reached
in the E- and P-criterion can only be judged by future
applications of these to real data.

\section*{Acknowledgments}

\indent

We thank Yaning Yang, Stephan Beirer, Hanspeter Herzel, Dirk 
Holste, Armin Schmitt, and Stefan Posch  for valuable discussions,
Wei Pan for recommending references, and NIH (N01-AR12256), 
NSF (0241102), and BMBF for financial support.

\newpage

\begin{figure*}[]
\centering
\begin{turn}{-90}
\epsfig{file=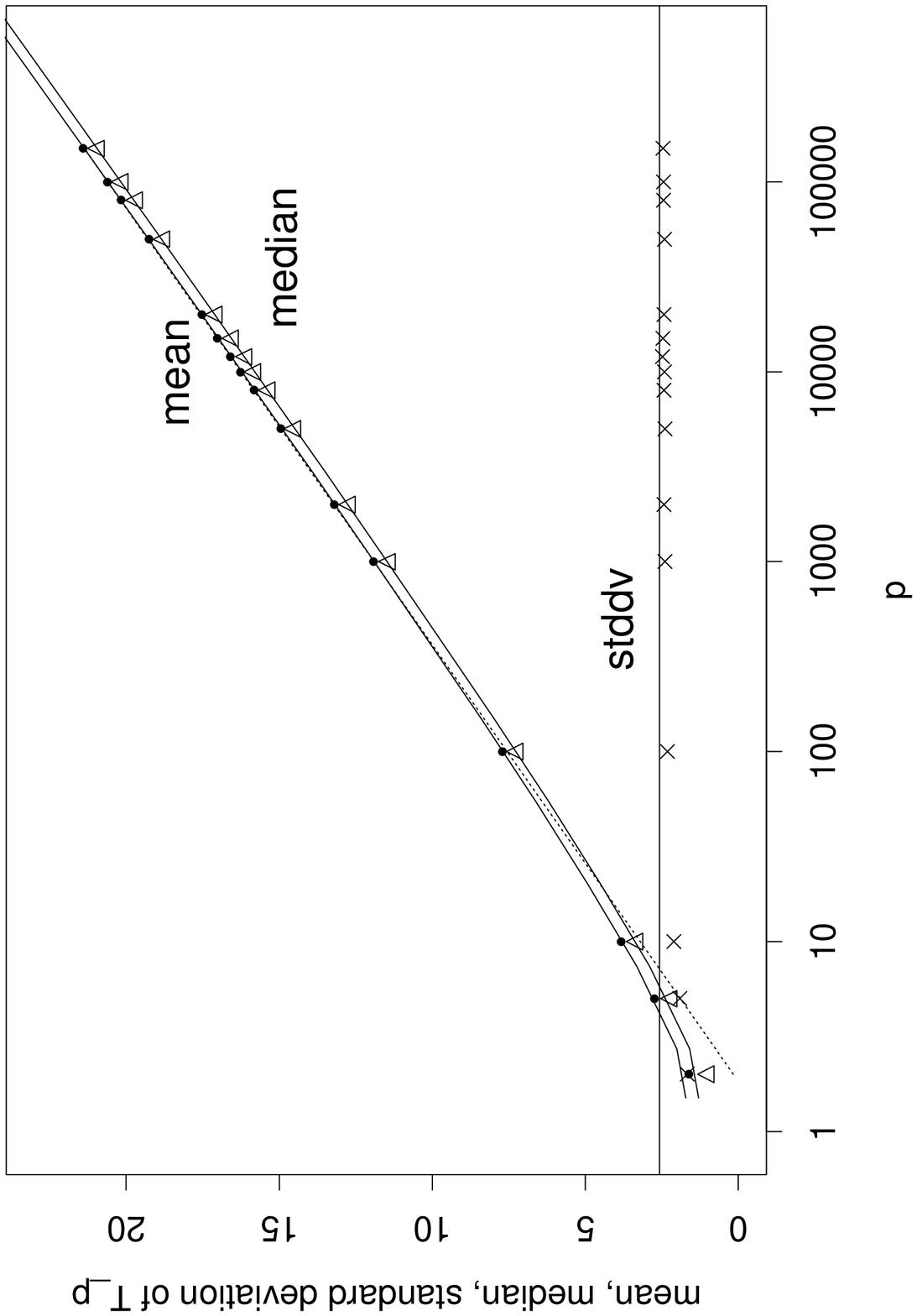, height=16cm}
\end{turn}
\caption{
Numerical simulation of the extreme-values $T_p = \max [t_1, t_2,
\ldots t_p]$ of $p$ random variables $t_1, t_2, \ldots, t_p$
sampled from the $\chi^2$ distribution with 1 degree of freedom.
The mean $E[T_p]$ (solid dots), the median $m[T_p]$ (triangles), and
the standard deviation $\sigma[T_p]$ (crosses) are plotted against
$\log (p)$ for $p$ ranging from $1$ to $1.5 \times 10^5$. The
analytic results of the mean, median, and standard deviation
by Eq.(\ref{eq_mean}), which are exact for asymptotic $p$,  are
shown in solid lines. For  asymptotically large $p$, both the mean and the
median of $T_p$ increase
with $p$ as $\sim 2 \log(p) - \log (\log(p))$.  A linear regression
line fitting the mean of $T_p$ is displayed in dashed line:
$E[T_p] \approx -1.14 + 1.89 \log (p)$
(the fitting range of $p$ is from $10^3$ to $1.5 \times 10^5$).
The horizontal solid line is the standard deviation
of $\sqrt{2 \pi^2/3} \approx 2.56$.
}
\end{figure*}

\begin{figure*}[]
\centering
\begin{turn}{-90}
\epsfig{file=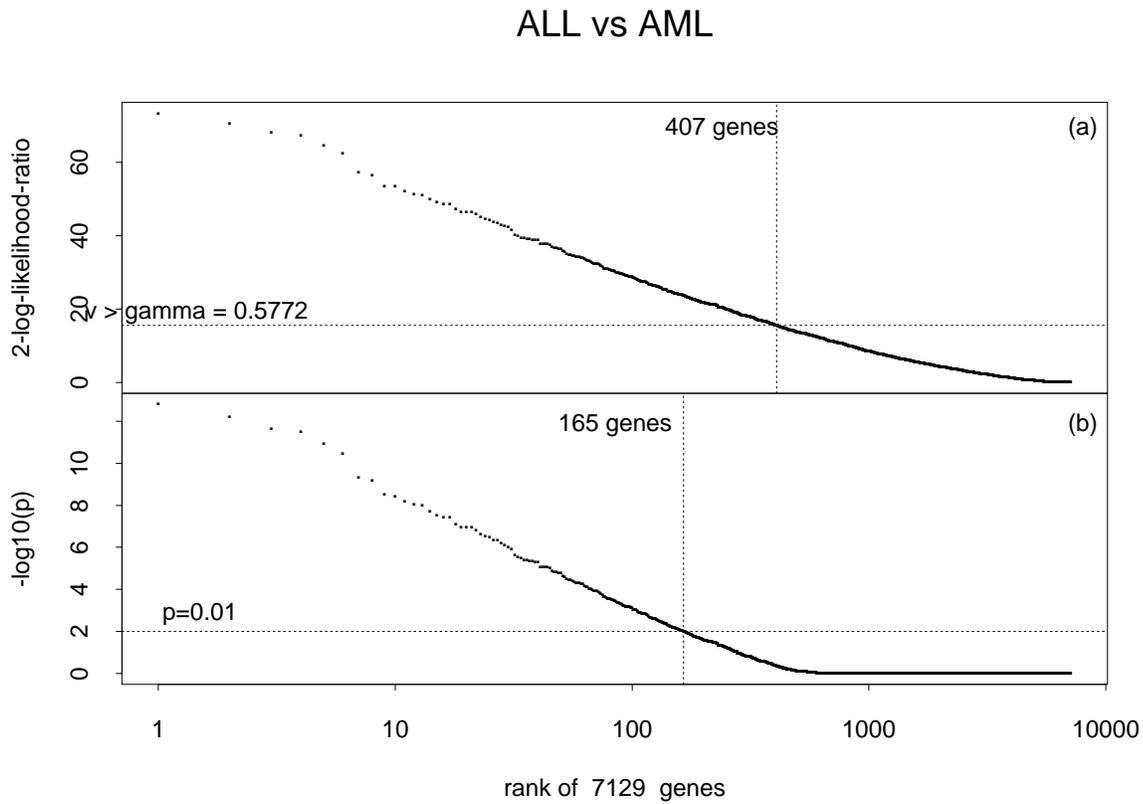, height=15cm}
\end{turn}
\caption{
(a) Rank-ordered log-likelihood ratios for ALL versus AML dataset:
$t_{(j)}$ ($j=1,2, \ldots p$) defined in Eq.(\ref{eq_tj}).
(b) Rank-ordered P-values for the same dataset:
$P_{(j)}= 1- \exp(- \exp(-v_{(j)}))$, where $v_{(j)}= (t_{(j)}- c_p)/2$.
In (a) E-criterion declares 407 genes as differentially expressed, 
and in (b) the more conservative P-criterion declares 165 genes 
as differentially expressed.  
}
\end{figure*}

\begin{figure*}[]
\centering
\begin{turn}{-90}
\epsfig{file=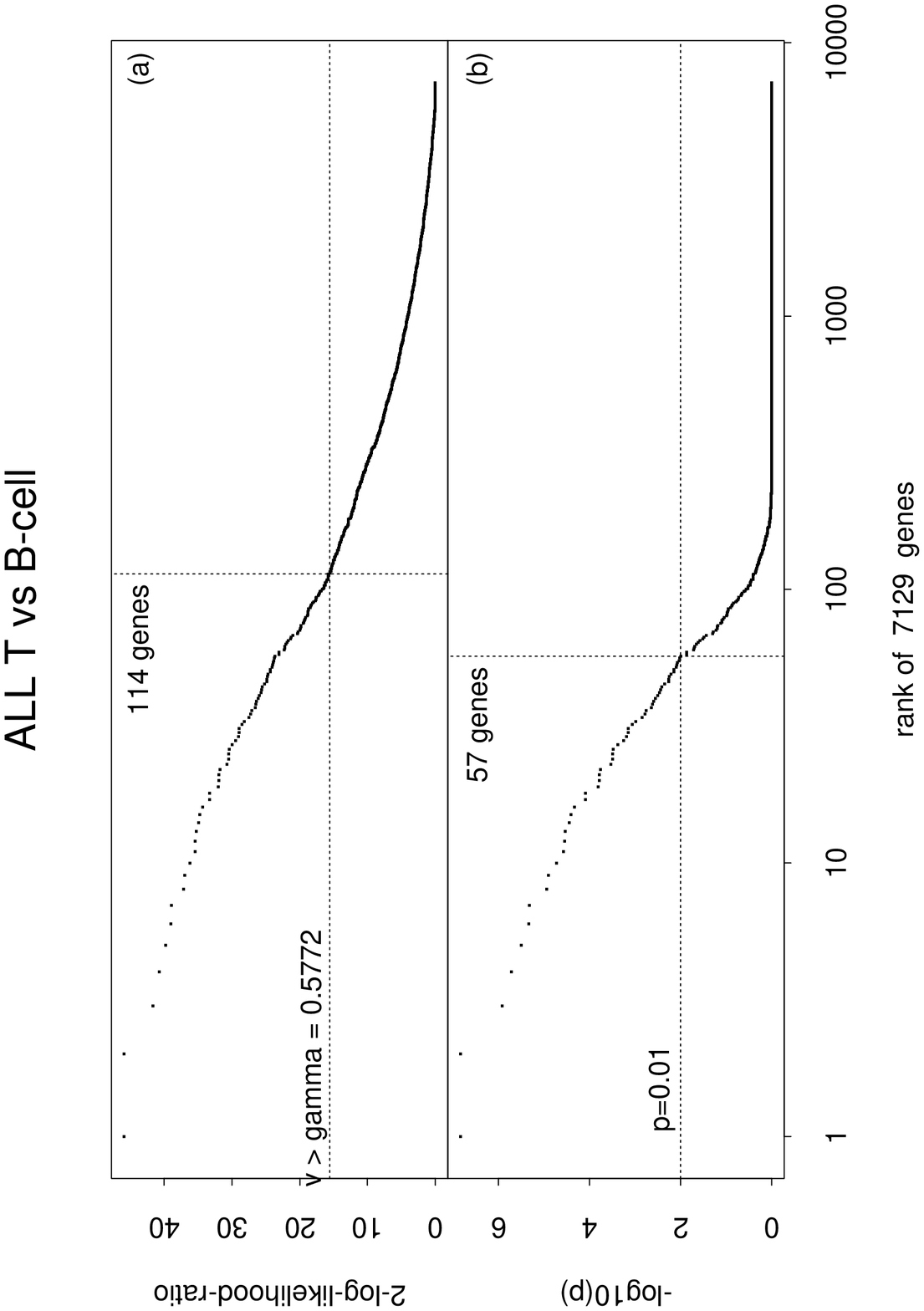, height=16cm}
\end{turn}
\caption{
(a) Rank-ordered log-likelihood ratios for  T-cell versus B-cell
dataset. (b) Rank-ordered P-values for the same dataset.
In (a) the E-criterion declares 114 genes as differentially
expressed, and in (b) the more conservative P-criterion
declares 57 genes as differentially expressed.
}
\end{figure*}

\begin{figure*}[]
\centering
\begin{turn}{-90}
\epsfig{file=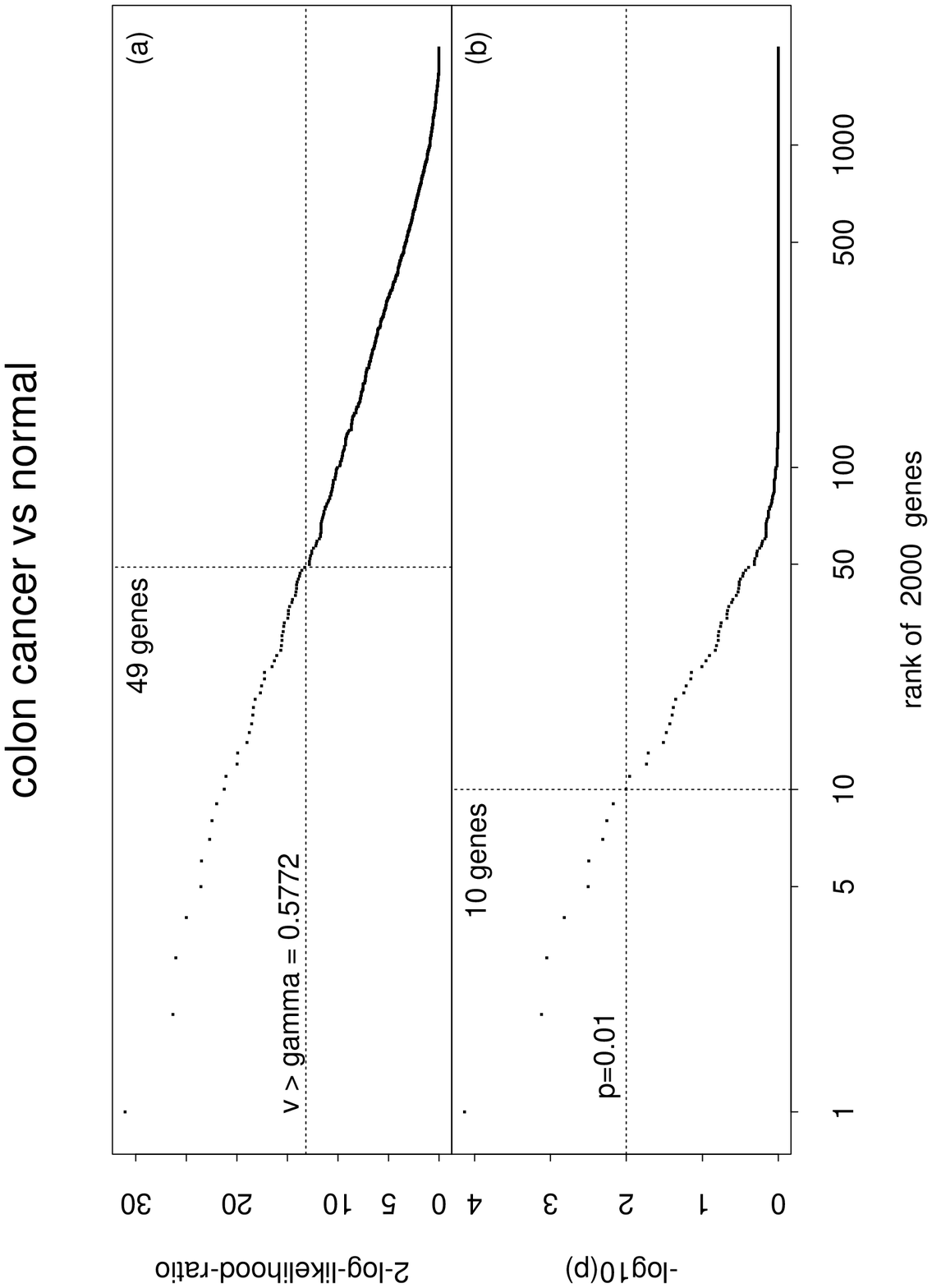, height=16cm}
\end{turn}
\caption{
(a) Rank-ordered log-likelihood ratios for colon versus normal
dataset.  (b) Rank-ordered P-values for the same dataset.
In (a) the E-criterion declares 49 genes as differentially expressed, 
and in (b) the more conservative P-criterion declares 10 genes as 
differentially expressed.
}
\end{figure*}

\end{document}